\documentclass[10pt,letterpaper]{article}
\usepackage{opex3}
\usepackage{amsmath}
\usepackage{amssymb}
\usepackage{cite}

\begin{document}

\title{Photovoltaic effect of light  carrying orbital angular momentum on  a  semiconducting stripe}

\author{J. W\"{a}tzel, A.~S. Moskalenko,$^*$ and J. Berakdar}

\address{Institut f\"{u}r Physik, Martin-Luther-Universit\"{a}t Halle-Wittenberg,
Heinrich-Damerow-Str. 4, 06120 Halle (Saale), Germany}

\email{$^*$ andrey.moskalenko@physik.uni-halle.de}



\begin{abstract}
We investigate the influence of a light beam carrying an orbital
angular momentum  on the current density of
an electron wave packet in a semiconductor stripe. It is
shown that due to the photo-induced torque  the electron
density can be deflected to one of the stripe sides. The direction of the
deflection is controlled by the direction of the light orbital momentum. In addition the net current density
can be enhanced.
This is a photovoltaic effect that
 can be registered by measuring the generated voltage drop across the stripe and/or the current increase.
\end{abstract}



\section{Introduction}

The possibility to produce light carrying an orbital angular momentum opened new opportunities in photonics
\cite{Molina2007,Allen_book}.  Such light, so-called twisted light (TL),
can be used to trap, rotate and manipulate microscopic particles
(optical tweezers) \cite{Allen2002,Barreiro2003,Friese1998}, atoms
and molecules \cite{Romero2002,Al-Awfi2000, Araoka2005}, as well
as Bose-Einstein condensates \cite{Helmerson_Torres2011twisted}.
TL can be used to generate electric currents in quantum
rings and hence a corresponding local light-controlled magnetic
fields \cite{Quinteiro2009}. The absorbtion of TL
by bulk semiconductors and the generation of local currents was
also demonstrated theoretically \cite{Quinteiro2009_EPL}. Further
investigations of the influence of TL on the
transport properties of charge carriers in semiconductors might be
interesting, e.g., in view of potential photovoltaic applications.
Currently, intensive research  are devoted to the improvement of the
efficiency of the  photovoltaic elements by adjusting their
material and structure properties;
Refs.~\cite{Gratzel2001,Peet2007,Thompson2008,Yang2005,Li2005,Timmerman2008}
are just but few  examples. An alternative approach might be based on
the adjustment of the structural properties of light before it is
used to create currents in the cells, e.g. employing TL which is the idea followed in this work.
 TL is created routinely from  usual light sources
  \cite{Beijersbergen1994,Heckenberg1992,Kennedy2002,Allen1992,Beijersbergen1993}, e.g.
 via traversing a spiral wave plate that can be deposited onto the solar cell. One may  think of enhancing the TL light intensity
via self-focusing effect, an issue addressed recently
 \cite{Thakur2010}. Strongly-focussed TL beams
deliver additional opportunities to manipulate optical properties
of nanostructures \cite{Quinteiro2009_PRB}. In this context new
horizons have just opened up due to the development of novel
optical lenses \cite{Zhang2008,Zhao2012} based on
metamaterials with a negative refractive index \cite{Shalaev2007}.

To explore the potential of TL for photovoltaics we  investigate how the focussed TL beam influences
an electron wave packet in a two-dimensional semiconductor stripe. We show that the
application of a TL beam results effectively in a voltage drop across the stripe with its sign
 being determined by the sign of the topological charge of TL.
 Our numerical calculations demonstrate this photovoltaic effect in the case of a GaAs-based semiconductor stripe irradiated by
 focussed TL beam that is generated  from the incoherent sunlight.

\section{Theoretical formulation}

\begin{figure}[t]
\centering\includegraphics[width=11cm]{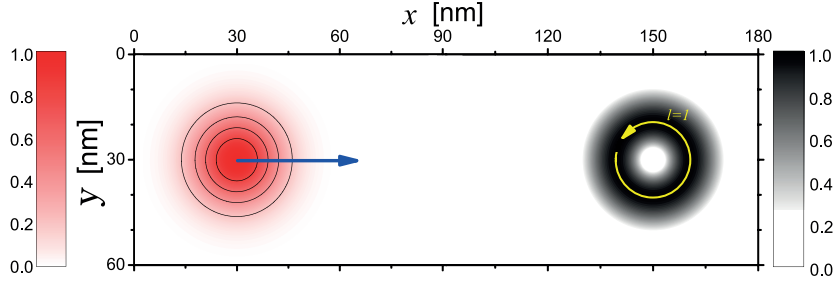}
\caption{Illustration of the system setup. In the left part the
normalized probability density corresponding to the initial wave function is
depicted. In the right part the normalized absolute value of the vector
potential $A(x,y,z=0,t=0)$ of the applied field is presented. The
electron has an initial momentum $k_x$ and will propagate in the
$x$-direction during the course of time. \label{system}}
\end{figure}

We consider an electron with an effective mass $m^{*}$  confined to a
two-dimensional stripe located in the plane $z=0$ (see
Fig.~\ref{system}). The width of the stripe in the $y$-direction
is denoted as  $L_0$ while the length in the $x$-direction is much
larger.  Nowadays such systems are routinely fabricated.
%
The electron motion is
driven by the incident electromagnetic field propagating
perpendicular to the stripe plane and is determined by the
time-dependent Schr\"{o}dinger equation:
\begin{equation}\label{Eq:TDSE}
    i\hslash\partial_t \psi=\left[-\frac{\hslash^2}{2m^{*}}
    (\partial_x^2+\partial_y^2)+\frac{ie\hslash}{m^{*}}(A_x\partial_x+A_y\partial_y)+\frac{e^2}{2m^{*}} A^2\right]
    \psi,
\end{equation}
where $\psi=\psi(x,y,t)$ is the wave function of the electron and
$\vec{A}=\vec{A}(x,y,z=0,t)$ is the vector potential of the
incident light in the plane of the stripe. Here we used
$\nabla \vec{A}=0$. We note however that in general the TL vector potential
is not completely transversal. The longitudinal component  is of  the order of
the wave vector of the light along the $z$ direction and hence, for  the frequencies considered here, can be  neglected
leading to the form of Eq.(\ref{Eq:TDSE}).
We assume that the initial state of the electron, before it enters
the area where it is influenced by the TL, can be
described by
\begin{equation}
\psi(t=0)=N\sin\left(\frac{\pi
y}{L_{_0}}\right)\exp\left[-\frac{(x-x_{_0})^2}{2\sigma^2}\right]\exp(i
k_x x)
\end{equation}
that corresponds to a freely propagating electronic wave packet in
the $x$-direction and the ground state for the motion in the
$y$-direction. Here $N$ is the normalization constant, wave vector
$k_x$ reflects the momentum of the propagating electron,
$(L_0/2,x_0)$ is the average electron position at $t=0$, and
$\sigma$ determines the width of the wave packet in the
$x$-direction.




In the course of time the electron reaches the region where it is
exposed to the influence of the TL beam. It is
important to note that we consider here the light beam focussed to
a very small spot beyond the diffraction limit. Such a focussing is
feasible thanks   to the recent achievements in the development of
novel lenses  based on metamaterials \cite{Zhang2008,Zhao2012}.
The profile of the electric field in such a spot is modified by
the lens arrangement with respect to the incident beam but its
topological structure is conserved. In the plane $z=0$, we model
the field of the TL beam carrying an orbital angular
momentum $l$ using the Laguerre-Gaussian (LG) modes
\cite{Allen1992} with an on-axis phase singularity of strength
$l$. This type of intensity distribution is also called optical
vortices \cite{Friese1998,ONeil2002,Simpson1997,Gahagan1996}.
Furthermore, the LG modes are characterized by the radial index
$p$ and the waist size $w_{0}$. In our case we use LG modes with
$p=0$ and $l\neq0$. In this case the field intensity profile in
the plane of the stripe  consists of a ring with a vanishing
intensity at the ring center (see Fig.~\ref{system}). The
corresponding vector potential of the circular polarized TL is convenient to write in polar
coordinates $(\tilde{r},\tilde{\phi})$ with the zero of this
coordinate system located at the spot center $(L_0/2,x_{\rm s})$.
It is given then by \cite{Allen1992}
\begin{equation}
\vec{A}(\tilde{r},\tilde{\phi},t)
={\rm
Re}\left\{\vec{e}A_0\left(\frac{\sqrt{2}\tilde{r}}{w_0}\right)^{l}\exp\left(-\frac{\tilde{r}^2}{w_0^2}\right)\exp\left[i(
l\tilde{\phi}-\omega t+\theta)\right]\right\},
\end{equation}
where  $\omega$ is the light frequency, $l$ is called the winding
number, $A_0$ characterizes the amplitude of the vector potential,
$\vec{e}$ is the polarization vector and $\theta$ is the phase.
We consider here the case of a circular polarization, i.e.
$\vec{e}=\vec{e}_+=\sqrt{1/2}(\vec{e}_x+i\vec{e}_y)$. The structure
of the LG beams is such that they can transfer an
orbital momentum and as a consequence a torque to an electron
\cite{Friese1998,ONeil2002,Simpson1997,Gahagan1996}. The amount of the transferred torque is
governed by  $l$. A practical advantage in this aspect is that the light-matter interaction is
much faster that the decay of the optical vortices. \\
Even though this condition
can be relaxed in principle, here we assume that during the electron motion across the considered
area with the TL spot impurity  scattering is
neglected. This can be assured by selecting a correspondingly  pure
semiconductor material.

An important aspect in our model is the degree of coherence of the
incident TL. The degree of first-order temporal
coherence of a stationary light source is determined via the
normalized first-order correlation function
\cite{Loudon2000quantum}
\begin{equation}
g^{(1)}(\tau)=\frac{\langle E^*(t) E(t+\tau)  \rangle}{\langle
E^*(t) E(t)  \rangle}, \label{firstordercoh}
\end{equation}
where the electric field is given by
$\vec{E}=-\partial_{t}\vec{A}$, which defines the coherence time of
light $\tau_{\rm c}$. In general, for $\tau\ll\tau_c$ the absolute
value of $g^{(1)}$ approaches 1 that means full coherence at this
time scale. For $\tau\gg\tau_c$ the absolute value of $g^{(1)}$ is
vanishing and the light is incoherent. We assume that the
propagation time $\tau$ of the electron through the TL
spot is much shorter than the coherence time of the light source
$\tau_{\rm c}$. Thus modelling the single electron motion we can
assume the light to be fully coherent on this scale and the phase
$\theta$ to be fixed in the corresponding single event. In order
to describe an average effect of the incoherent light beam on many
propagating electrons in a time period much larger than $\tau_{\rm
c}$ we have to assume the phase $\theta$ to be random within the
interval $[0,2\pi)$. Any calculated physical property has then to
be averaged over the random phase distribution. In practice, we
have to simulate a large enough number $N$ of realizations using
the random number generator for the phase and then average the
calculated physical quantity over these realizations
\cite{Loudon2000quantum}.

\section{Numerical results}

In our numerical simulations we consider a stripe made of GaAs with a width of $L_{_{0}}=60$ nm. The simulated length of the stripe in the $x$-direction is taken to be $400$~nm to avoid any reflections of the wave function in this direction by the barriers. Figure \ref{system} shows $180$~nm from the whole length in the $x$-direction.
The electron has an effective mass $m^{*}=0.067m_{_0}$, where $m_{_0}$ is the free electron mass. The parameter $\sigma$, which characterizes the width of the density of the electron in the $x$-direction, is $15$ nm, whereas the width in the $y$-direction is determined by the properties of the ground state. The initial velocity of the electron is $7.25\times10^6$ m/s, which is achieved by accelerating the electron with a voltage of $10$ V. The waist of the beam $w_{_{0}}$ is taken to be 12.5 nm, that means the TL spot is focussed within the stripe. The center of the spot is localized at (150\,nm, 30\,nm), that means the  center-center distance between the initial density of the electron and the vector potential is $120$ nm. Consequently, the electron needs a time of $\tau=24.83$ fs to propagate completely from the initial position through the TL spot. The spectral width of the natural sun light is $\Delta\nu\approx5\times10^{14}$ Hz, i.e. the coherence time is given by $\tau_{\rm c}=1/\Delta\nu\approx2$ fs. This small value shows that unfiltered sun light is not appropriate for our simulations because the propagation time $\tau$ of the electron is much larger than the coherence time of light, i.e. the TL is incoherent at this time scale. To overcome this limitation one can restrict the spectrum of the natural light to one color, with the help of an appropriate spectral filter. In case of yellow the spectral width is $18\times10^{12}$ Hz, which leads to a coherence time $\tau_{\rm c}=55.55$ fs. Consequently, $\tau<\tau_{\rm c}$ and therewith we can assume that the TL spot remains coherent during the electron propagation through the TL spot. Our intention is that the TL spot should be produced from the incident sunlight. The intensity of the solar electromagnetic radiation
at the top of the atmosphere is $I=1.361$ kW/m$^2$ \cite{solarconstant}.  This value is weakened due to the propagation through the atmosphere, spectral filtering and focussing.
In our simulations we use therefore the value $I=500$ W/m$^2$ for the peak intensity. This corresponds to the amplitude of the electric field $E_{_0}=\sqrt{\frac{2I}{\epsilon_{0}c}}=613.4$ V/m.

\begin{figure}[t]
\centering\includegraphics[width=11cm]{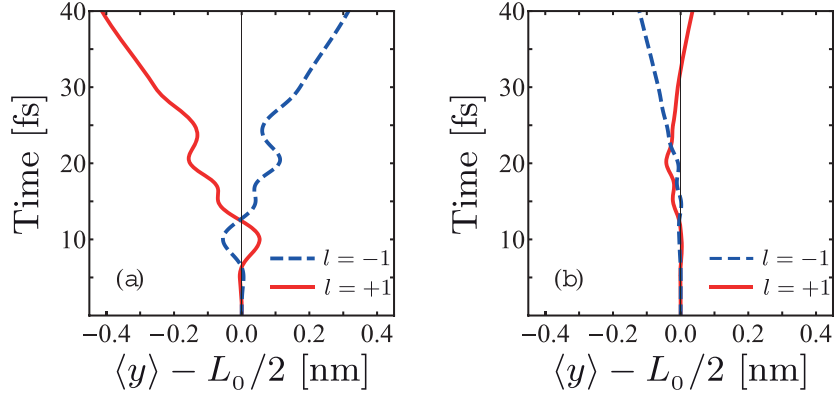}
\caption{Calculated  time-dependent expectation value $\langle y\rangle(t)$ in the case of (a) a TL spot with the fixed phase $\theta=0$ and (b) the averaged effect of 10000 electrons with random phases $\theta\in[0,2\pi)$. Once the electron wave reaches the TL spot, it begins to drift in the $y$-direction. This drift depends on the sign of the orbital momentum $l$.}
\label{ExpV}
\end{figure}

Let us consider the effect of the TL on the moving electron under such conditions. In Fig. \ref{ExpV} the time-dependent expectation value of the electron position in the $y$-direction $\langle y\rangle(t)$ is plotted for two situations. Panel (a) shows $\langle y\rangle(t)$ in case of a simulation of the motion of the electron through a TL spot with a fixed phase $\theta=0$. Panel (b)
shows the averaged position $\langle y\rangle(t)$ in case of 10000 simulations, whereas every TL spot has a random phase $\theta\in[0,2\pi)$. In both cases we consider two different directions of the angular momentum of the TL $l=1$ and $l=-1$, respectively.
We infer that depending on the sign of the orbital angular momentum    $l$,
 the deviation from the initial position at $y=L_{_0}/2$ is directed to the left or to the right side of the stripe.
In the case of the averaged effect of 10000 realizations with random phases $\theta\in[0,2\pi)$ of the TL spot we also obtain a deviation which is on the same order of magnitude as in the case of a single realization with a fixed phase $\theta$. This is a remarkable result because it shows that the TL spot can deflect the moving electron stream to one or another side of the stripe in dependence on the selected angular momentum sign. That is not possible for a conventional light spot without angular momentum (e.g., a Gaussian beam). The time dependence of the position deviation becomes almost linear for $t>25$ fs, after the full wave packet passed the TL spot. In the case of a single realization with a fixed phase $\theta=0$, the deviation is directed to the left (right) side of the stripe for $l=+1$ ($l=-1$), while in the case of 10000 realizations with random phases it is directed to the right (left) side of the stripe.
Notice also that the expectation values corresponding to $l=\pm1$ do not behave symmetrically with respect to the middle of the stripe.
Particularly, in the case of the electron stream this asymmetry is pronounced. These apparently contradictious observations can be explained by recalling that the applied light is circularly polarized. During the electron propagation through the TL spot the direction of the light polarization, and therefore the direction of the force acting on the electron, changes leading to the oscillations in the time dependence of the averaged position observed in Fig.~\ref{ExpV}(a). There are only several oscillation cycles performed by the field before the electron leaves the interaction area. Therefore, the final result depends strongly on the phase of the electromagnetic field and this effect overlaps the effect arising because of the orbital momentum of the TL. Deflections to the right side and to the left side can be observed for both values of $l$ depending on the phase of the field. However, for an electron stream they
average to zero in the case of a conventional light spot and lead to the pure effect of the orbital momentum in the case of a TL spot, shown Fig.~\ref{ExpV}(b).
\begin{figure}[t]
\centering\includegraphics[width=11cm]{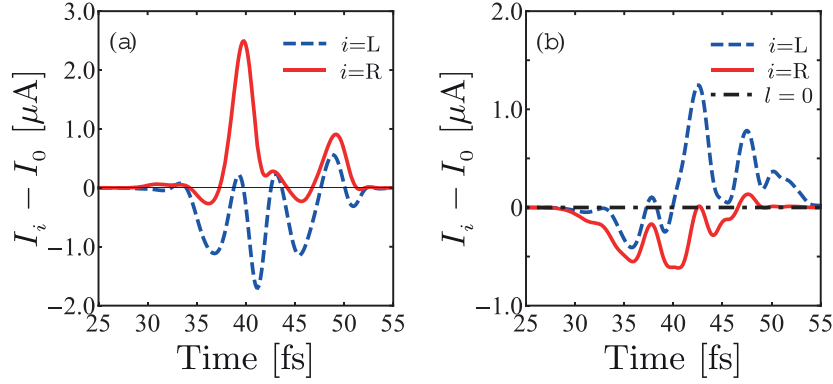}
\caption{Time-dependent difference between the side currents $I_{_i}$ ($i$=L,R) and the corresponding currents $I_{_0}$  generated by a freely moving electron (i.e. $A=0$) in case of $l=-1$ is shown, whereas the situation with (a) a fixed phase $\theta=0$  and  (b) the averaged effect of 10000 realization with random phases $\theta$ can be seen. Furthermore the averaged effect in case of light without angular momentum (i.e. $l=0$) is shown.
}
\label{current}
\end{figure}
In order to observe the symmetry between the cases of $l=+1$ and $l=-1$, one must also change the direction of the circular polarization $(\vec{e}_{_+}~\rightarrow~ \vec{e}_{_-})$. This means that e.g. for $l=+1$ and the left-circularly polarized TL ($\vec{e}=\vec{e}_{_+}$)
one obtains a deflection being mirror-symmetric to the case of $l=-1$ and the right-circularly polarized light ($\vec{e}=\vec{e}_{_-}$).

Another interesting aspect is that the structure of the time-dependent current density. Due to the symmetry of the problem it is reasonable to introduce the current flowing through the left side of the stripe that is given by
\begin{equation}
I_{_{\rm L}}(t)=\int_{0}^{L_{_0}/2}{\rm d}y\,j_{_{x}}(x_{_{\rm d}},y,t)\;,
\end{equation}
where $j_{_{x}}(x,y,t)=\frac{i\hslash}{2m^{*}}\left[\psi\partial_{_x}\psi^{*}-\psi^{*}\partial_{_x}\psi\right]$ is the time-dependent probability current density in the $x$-direction.
The current flowing through the right side is given by an analogous equation with the range of integration from $L_{_0}/2$ to $L_{_0}$. These currents in the $x$-direction would be detected along a line in the $y$-direction at a position $x=x_{_{\rm d}}$ behind the TL spot.

In Fig. \ref{current} the difference between the generated side currents $I_{_i} (i={\rm L,R})$ and the side currents $I_{_0}$ corresponding to
a freely moving electron (i.e. $A=0$), which are equal for both sides, are illustrated. We show the results in case of the quantum number $l=-1$ of the TL spot. The hypothetical detector is positioned at $x_{_{\rm d}}=180$ nm.
The situation with one fixed phase $\theta=0$ and the averaged effect for random phases, corresponding to the case of an electron stream, are shown. It is the averaged effect that is of interest for a practical realization. One can see that the difference $I_{_{\rm L}}-I_{_0}$ is positive for nearly the whole propagation time through the detector, while in case of the right side current the difference is mostly negative. Thus we have found an increase of the current on the left side relative to the case of a freely moving electron, i.e. without any external field. Simultaneously the current on the right side of the stripe decreases.
These current changes can be explained by the deflections of the electronic density illustrated in Fig.~\ref{ExpV}. For example, we see
from Fig.~\ref{ExpV}(b) that in case of $l=-1$ the electron stream is deflected to the left side.
It means that there is more electron density flowing on the left side, i.e. the probability current is larger than on the right side, that is observed in Fig.~\ref{current}(b).

\begin{figure}[t]
\centering\includegraphics[width=11cm]{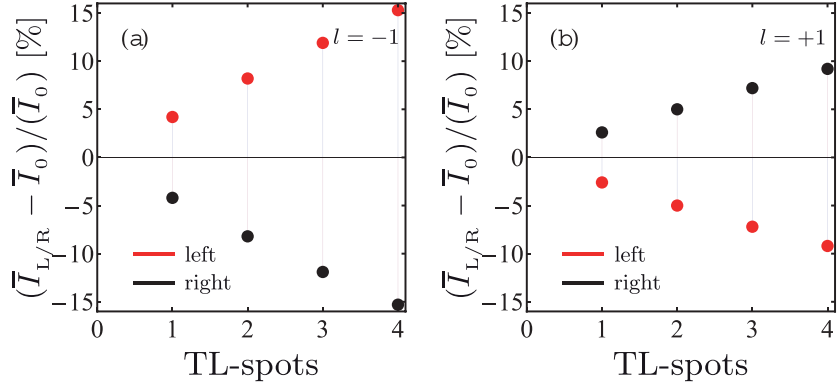}
\caption{Relative change of the currents on the left and right sides in dependence on the number of TL spots. The cases for (a) $l=-1$  and (b) $l=+1$  are shown.}
\label{SevL}
\end{figure}

Averaged over the time profile we obtain an increase of the left side current $\overline{I_{_{\rm L}}}$ of the electron stream, influenced by the TL spot with $l=-1$, by 4.2\% relative to $\overline{I_{_0}}$. Correspondingly, the right side current $\overline{I_{_{\rm R}}}$ decreases by the same amount.
The physical reason for this effect is the torque transferred to the electron from the TL.
In case of $l=+1$, we obtain an increase (decrease) of the right (left) side current by 2.6\%.
We did not obtain the same values as in case of $l=-1$, because only the sign of the orbital angular momentum $l$ was changed. The polarization direction was not changed and consequently $I_{_{\rm L/R}}(l=+1)\neq I_{_{\rm R/L}}(l=-1)$. As it is mentioned above, the equality is achieved only by a simultaneous change of the sign of $l$ and the direction of the circular polarization. In case of light without orbital angular momentum (i.e. $l=0$) the averaging over 10000 realizations shows that one can not expect different side currents. In fact our numerical results reveal that the difference $I_{_{L/R, l=0}}-I_{_0}$ approaches zero for a large number of realizations, which is observed in Fig.~\ref{current}(b).\\
The question arises how the effect of current increase or decrease can be enhanced. To do this we placed additional light spots on the stripe. As a consequence the electron travels through several light spots which have the same properties and the same distance between each other. In Fig. \ref{SevL} one can see that the side current changes  depending  on the number of TL spots.
With every additional TL spot the increase/decrease of the corresponding side currents enhances. Just four TL spots are sufficient to reach an increase of 15\% in case of $l=-1$. In case of $l=+1$ we obtain an increase of 9\%.

\section{Conclusion}

We have shown that  light carrying the orbital angular momentum
can be utilized to generated and manipulated  charge currents flowing along  the
right and the left sides of a two-dimensional semiconductor
stripe. These changes have an opposite sign and can be controlled
by selecting the sign of the winding number of the beam. The effect can be
magnified by increasing the number of twisted light spots in the
stripe. With four twisted light spots we are able to demonstrate a
change of the corresponding side currents by 15\%.

\section*{Acknowledgment}
We acknowledge  helpful discussions with Dr. Guillermo
Quinteiro on the twisted light properties. J.W. was partly
supported by the Gustav Mie prize of the Martin Luther University
Halle-Witteberg. J.B. is supported by the DFG through SFB 762.

\end{document}